\documentclass[english,aps,twocolumn,showpacs,amsmath,amssymb,prb,superscriptaddress]{revtex4-1}
\usepackage[T1]{fontenc}
\usepackage[latin9]{inputenc}
\usepackage{color}
\usepackage{amstext}
\usepackage{amssymb}
\usepackage{graphicx}

\makeatletter

\@ifundefined{textcolor}{}
{%
 \definecolor{BLACK}{gray}{0}
 \definecolor{WHITE}{gray}{1}
 \definecolor{RED}{rgb}{1,0,0}
 \definecolor{GREEN}{rgb}{0,1,0}
 \definecolor{BLUE}{rgb}{0,0,1}
 \definecolor{CYAN}{cmyk}{1,0,0,0}
 \definecolor{MAGENTA}{cmyk}{0,1,0,0}
 \definecolor{YELLOW}{cmyk}{0,0,1,0}
}

\usepackage[caption=false]{subfig}

\@ifundefined{showcaptionsetup}{}{%
 \PassOptionsToPackage{caption=false}{subfig}}
\usepackage{subfig}
\makeatother

\usepackage{babel}
\begin{document}

\title{Origami rules for the construction of localized eigenstates of the
Hubbard model in decorated lattices}

\author{R. G. \surname{Dias}}

\email[To whom correspondence should be addressed: ]{rdias@ua.pt}

\affiliation{Department of Physics, I3N, \\
 University of Aveiro, \\
 Campus de Santiago, Portugal}

\author{J. D. \surname{Gouveia}}

\affiliation{Department of Physics, I3N, \\
 University of Aveiro, \\
 Campus de Santiago, Portugal}

\date{\today}
\begin{abstract}
We present a method of construction of exact localized many-body eigenstates
of the Hubbard model in decorated lattices, both for $U=0$ and $U\rightarrow\infty$.
These states are localized in what concerns both hole and particle
movement. The starting point of the method is the construction of
a plaquette or a set of plaquettes with a higher symmetry than that
of the whole lattice. Using a simple set of rules, the tight-binding
localized state in such a plaquette can be divided, folded and unfolded
to new plaquette geometries. This set of rules is also valid for the
construction of a localized state for one hole in the $U\rightarrow\infty$
limit of the same plaquette, assuming a spin configuration which is
a uniform linear combination of all possible permutations of the
set of spins in the plaquette.
\end{abstract}


\maketitle
The field of itinerant geometrically frustrated electronic systems
has attracted considerable interest in the last two decades \citep{Montenegro-Filho2006,Tasaki1998,Derzhko2005,Derzhko2010,Duan2001,Richter2004,Schulenburg2002,Mielke1992,Derzhko2004,Richter2004a,Montenegro-Filho2014,Gulacsi2007,Lopes2014,Rojas2012,Derzhko2015,Leykam2013,Tamura2002,Mielke1999,Wu2007,Mielke1993,Tanaka2007,Tasaki1992,Tasaki1994,Tasaki1998a,Tasaki1995,Mielke1991a}.
Much of this interest was related with the study of flat-band ferromagnetism
in these systems \citep{Tamura2002,Mielke1999,Wu2007,Mielke1993,Derzhko2015,Leykam2013}.
Flat-band ferromagnetism occurs in decorated lattices of the Mielke's
and Tasaki's classes, which display degenerate localized ground states
with overlapping probability densities \citep{Tanaka2007,Tasaki1992,Tasaki1994,Tasaki1995,Tasaki1998a,Mielke1991a}.
The emerging ferromagnetism can be interpreted as resulting from a
generalized Hund's rule \citep{Froehlich2005}. In the case of the
lattices which fall into the Lieb's class, the flat bands intercalate
itinerant bands \citep{Nita2013} and mean-field studies of the Hubbard
Hamiltonian in the Lieb lattice indicate that for large $U$, ferromagnetism
is expected except near half-filling where a ferrimagnetic phase appears
\citep{Gouveia2015,Lieb1989}.

These localized states are one-particle eigenstates of the tight-binding
Hamiltonians for the decorated lattices and little is known about
the many-body eigenstates of an interacting system of fermions in
decorated lattices \citep{Movilla2011} (assuming Hubbard-like interactions),
besides the appearance of a ferromagnetic ground state in decorated
lattices of the Mielke's and Tasaki's classes \citep{Mielke1992,Tasaki1992}.
In particular, the interacting ground state of the Hubbard model is
not known in the case of lattices of the Lieb's class. Approximate
analytic results can be obtained in the weak coupling limit, addressing
interactions as perturbations of the tight-binding flat bands \citep{Bodyfelt2014}.

In this manuscript, we present a method of construction of exact localized
many-body eigenstates of the Hubbard model in decorated lattices
of arbitrary dimensions,
for $U=0$ and $U\rightarrow\infty$. These states are localized in
what concerns hole and particle movement. This
method relies in simple arguments which lead to a set of quantum ``origami''
rules: i) if one plaquette or a set of plaquettes has a higher symmetry
than that of the whole lattice, one can find energy eigenstates that
have zero probability density at the sites that connect the plaquette
or the set of plaquettes to the rest of the lattice (this argument
is enough to justify the existence of localized states in the case
of
two-dimensional
decorated lattices of the Lieb's class); ii) given such a localized
state in the symmetric plaquette, one can fold the plaquette, either
at the probability density nodes or at other equivalent sites (adjusting
the probability density at those sites and the hopping constants that
involve those sites), therefore lowering the symmetry of the plaquette;
iii) the energy of the localized state can be lowered by adding hopping
terms between sites with the same localized state phase (if the hopping
constant is negative) or hopping terms between sites with opposite
phases (if the hopping constant is positive). Hopping terms between
nodes of the localized state may also be added, but do not change
the energy of the localized state. The hopping terms added must preserve
the symmetry of the localized state. These two arguments justify localized
states in decorated lattices of the Mielke's and Tasaki's classes.
Furthermore, the spin degree of freedom of the $U=0$ Hubbard  Hamiltonian may be interpreted as a  sublattice index and localized states can also be created using these origami rules involving the two (up and down spin) sublattices.
Such localized states arise for instance as edge states in 1D tight-binding descriptions of topological insulators  \cite{Shen2013,Qi2011,Hasan2010}.

The remaining part of this paper is organized in the following way.
First, we review the construction of one-particle localized eigenstates
of the tight-binding decorated lattices of the Lieb's class. We then
generalize this construction to more complex lattices using a symmetry
argument and introducing the set of origami rules. Next, we show how
to extend these rules to the case of the $U\rightarrow\infty$ limit
of the Hubbard model. Finally we conclude.

The Hubbard Hamiltonian in a decorated lattice can be written as
\begin{equation}
H=\sum_{\langle ij\rangle,\sigma}t_{ij}c_{i\sigma}^{\dagger}c_{j\sigma}+U\sum_{i}n_{i\uparrow}n_{i\downarrow},\label{eq:H1}
\end{equation}
where the creation (annihilation) of an electron at site $i$ with
spin $\sigma$ is denoted by $c_{i\sigma}^{\dagger}$ ($c_{i\sigma}$)
with $n_{i\sigma}$ being the number operator $n_{i\sigma}=c_{i\sigma}^{\dagger}c_{i\sigma}$
and $n_{i}=n_{i\uparrow}+n_{i\downarrow}$. The sum over $\langle ij\rangle$
is the sum over all pairs of sites with a finite hopping probability
between them and this a different sum for each decorated lattice.
The hopping constants are assumed to be equal, $t_{ij}=t$, unless
stated otherwise. When $t=0$, all states with the same number $N_{d}$
of doubly occupied sites are degenerate. In this paper, we assume
$N_{d}=0$. The Hubbard model in the limit $U\rightarrow\infty$ is
also designated as Harris-Lange model \citep{Harris1967}. In this
limit, using the identity $c_{i\sigma}=c_{i\sigma}[(1-n_{i\sigma})+n_{i\sigma}]$,
the Hubbard model can be rewritten as
\begin{equation}
\hat{H}=\sum_{\langle ij\rangle,\sigma}t_{ij}(1-n_{i\bar{\sigma}})c_{i\sigma}^{\dagger}c_{j\sigma}(1-n_{j\bar{\sigma}})+U\sum_{i}n_{i\uparrow}n_{i\downarrow}\label{eq:harris}
\end{equation}
with $\bar{\sigma}=-\sigma$. An important point about the strong
coupling limit is that the Hamiltonian eigenfunctions, in the case
of a Hubbard ring, can be written as a tensorial product of the eigenfunctions
of a tight-binding model of independent spinless fermions (holes)
in the ring with $L$ sites and the eigenfunctions of an Heisenberg
model (with exchange constant $J=t^{2}/U$) in a reduced chain \citep{Ogata1990,Gebhard1997,Dias1992,Peres2000}.

Let us first discuss the $U=0$ case. Flat bands in the one-particle
tight-binding energy dispersion of geometrically frustrated lattices
reflect the existence of degenerate localized eigenstates which are
translated versions of the same state $|\psi_{\text{loc}}\rangle$.
The probability density associated with one of these localized
states is non-zero only in a small lattice region. In the particular
case of decorated lattices of the Lieb's class, the localized states
can be viewed as one-dimensional standing waves in tight-binding rings,
associated with paths in the 2D lattice which include one or two plaquettes
\citep{Lopes2011}. For zero flux, all one-particle energy levels
of a tight-binding ring (with even number of sites) are doubly degenerate
(except for $k=0$ and $k=\pi$) and the respective eigenstates have
opposite momenta. Adding or subtracting the states of opposite momenta,
one obtains a standing wave with a number of nodes that depends on
$k$. If these nodes coincide with the sites at the vertices of a
plaquette of a decorated lattice, the electron becomes trapped inside
the plaquette. Therefore, flat band eigenstates of decorated lattices
of the Lieb's class are constructed from standing waves such that
the nodes coincide with sites at the vertices. Note that these localized
states overlap in real space, that is, they constitute a basis of
the subspace of localized states but not an orthogonal basis.

\begin{figure}
\includegraphics[width=0.9\columnwidth]{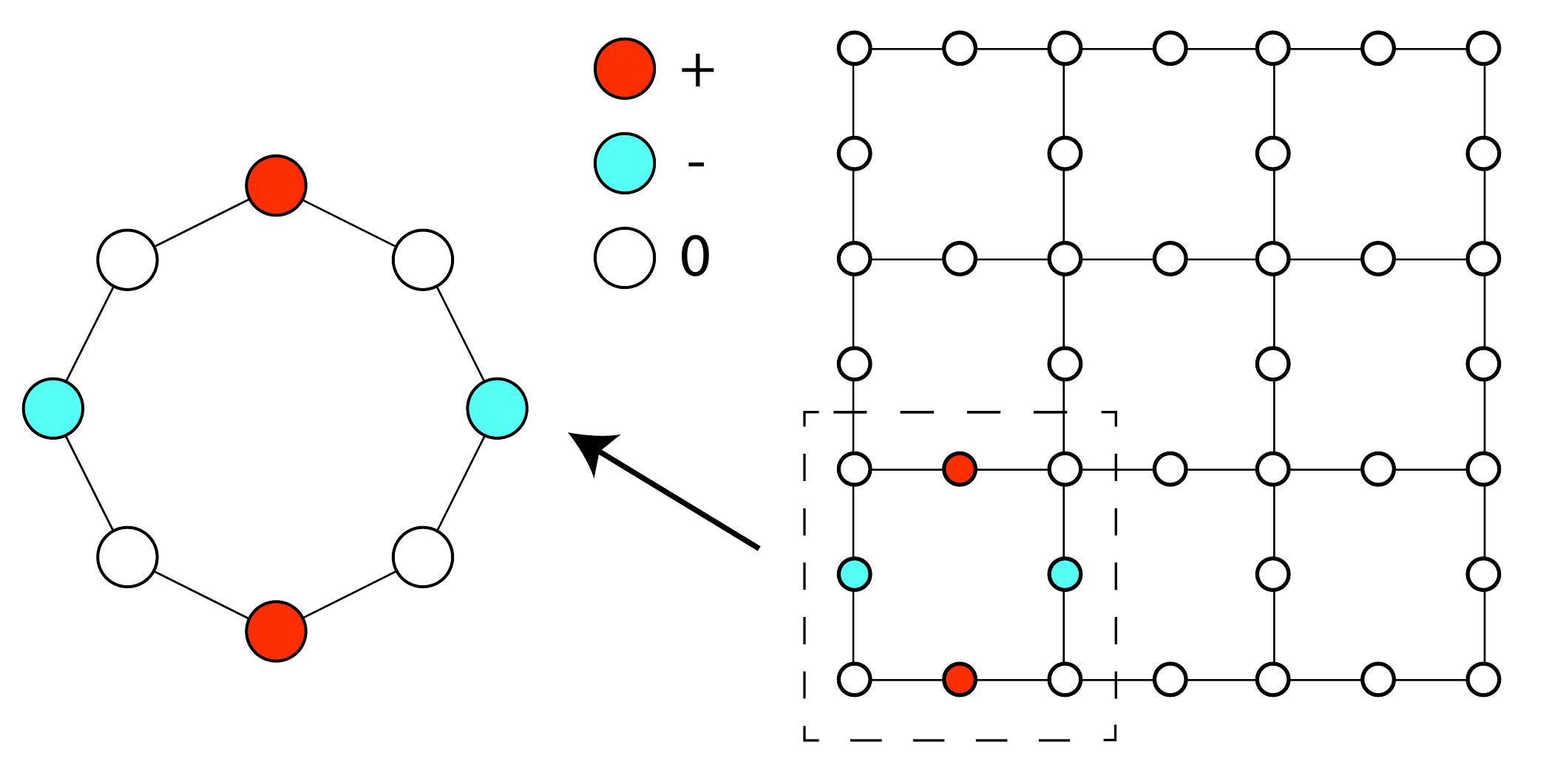}\protect\caption{The symmetry of the tight-binding Hamiltonian for the Lieb plaquette
is the same as that of the tight-binding ring and larger than that
of the Lieb lattice. The distance between adjacent sites is assumed
to be $a=1$. All hopping constants are equal. \label{fig:symmetry}}
\end{figure}

\begin{figure}
\subfloat[\label{fig:localizedouteronly}]{\includegraphics[width=2cm]{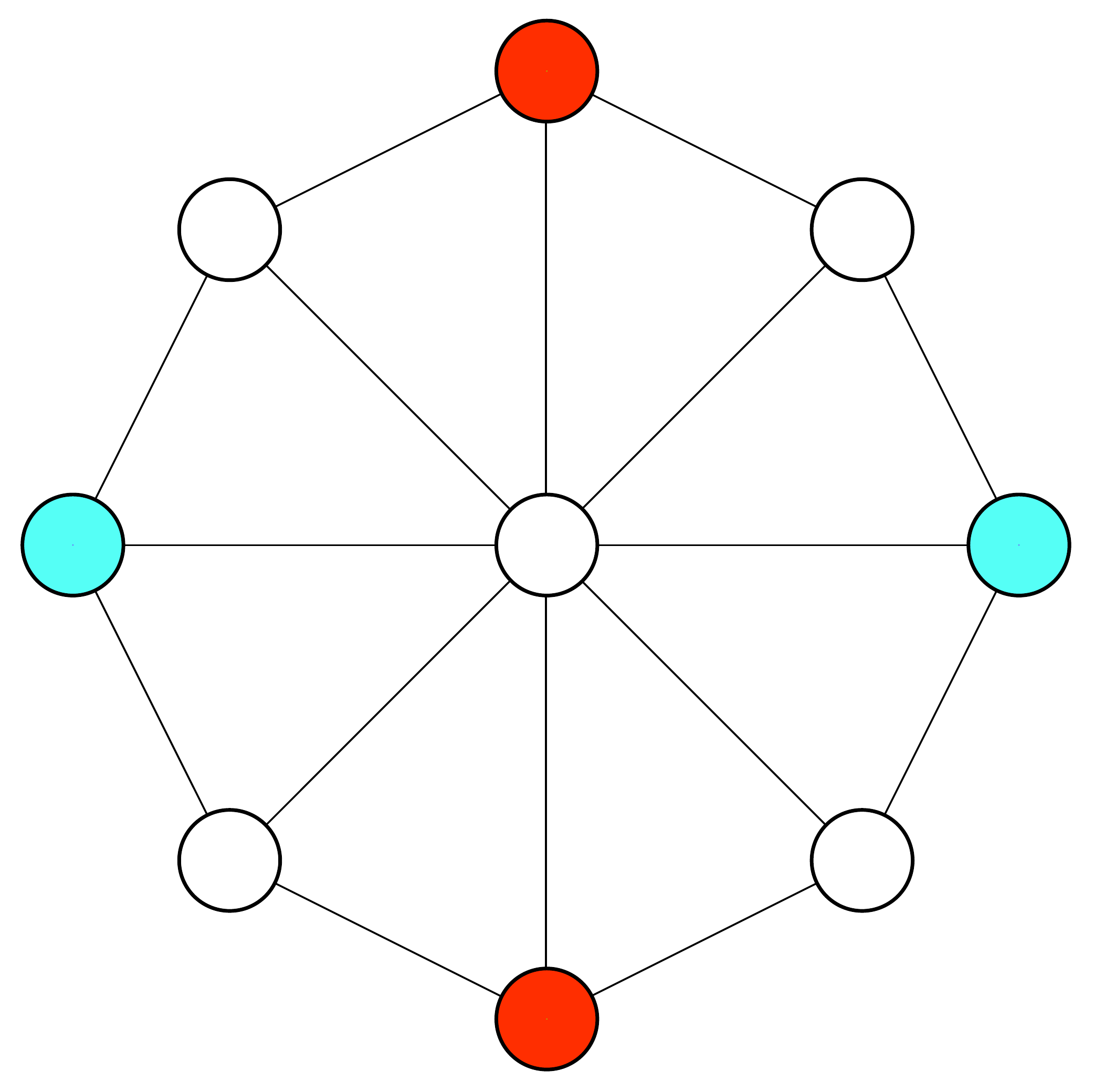}
}\subfloat[]{\includegraphics[width=2cm]{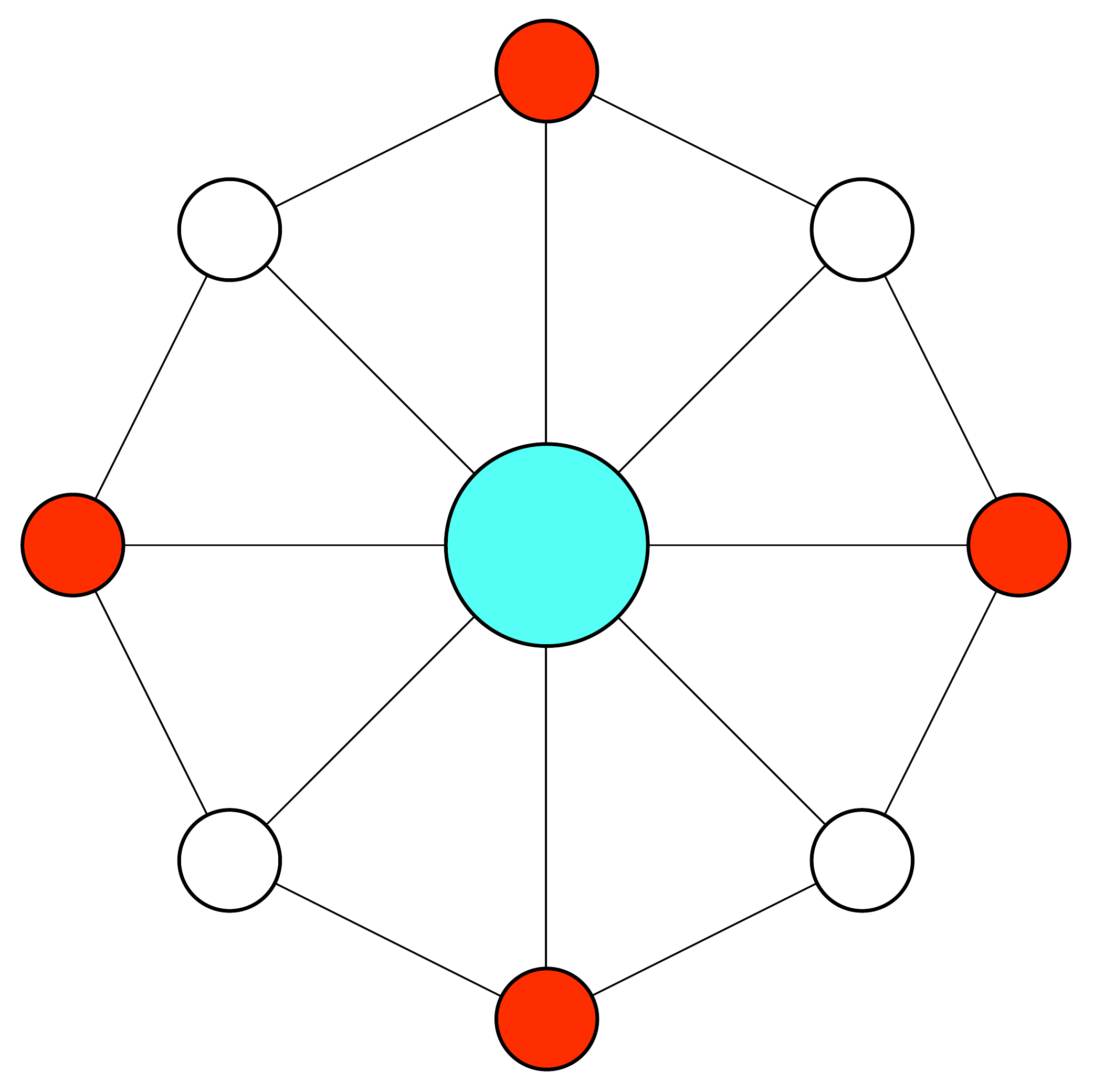}
}\subfloat[]{\includegraphics[width=2cm]{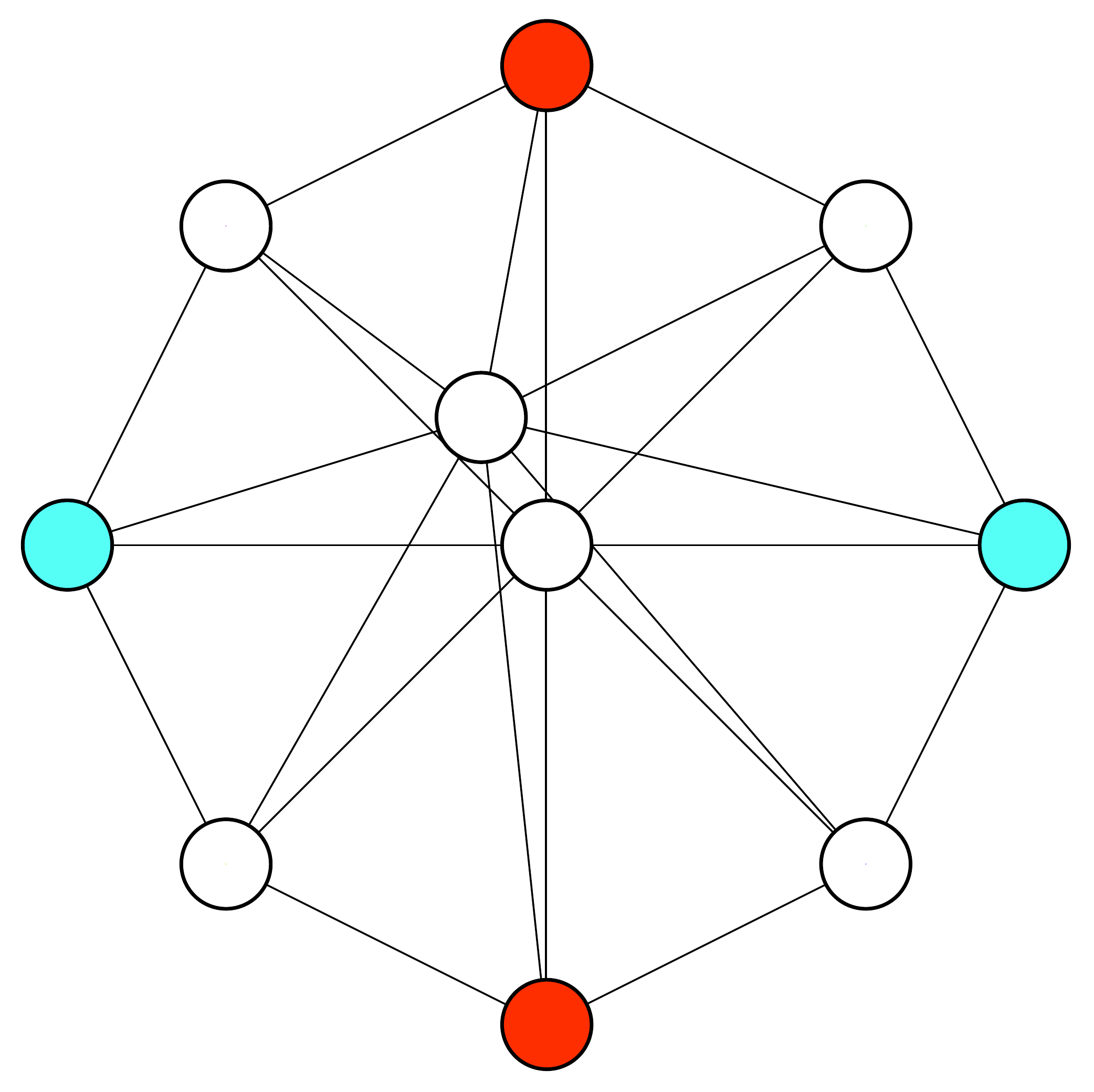}}
\subfloat[]{\includegraphics[width=2cm]{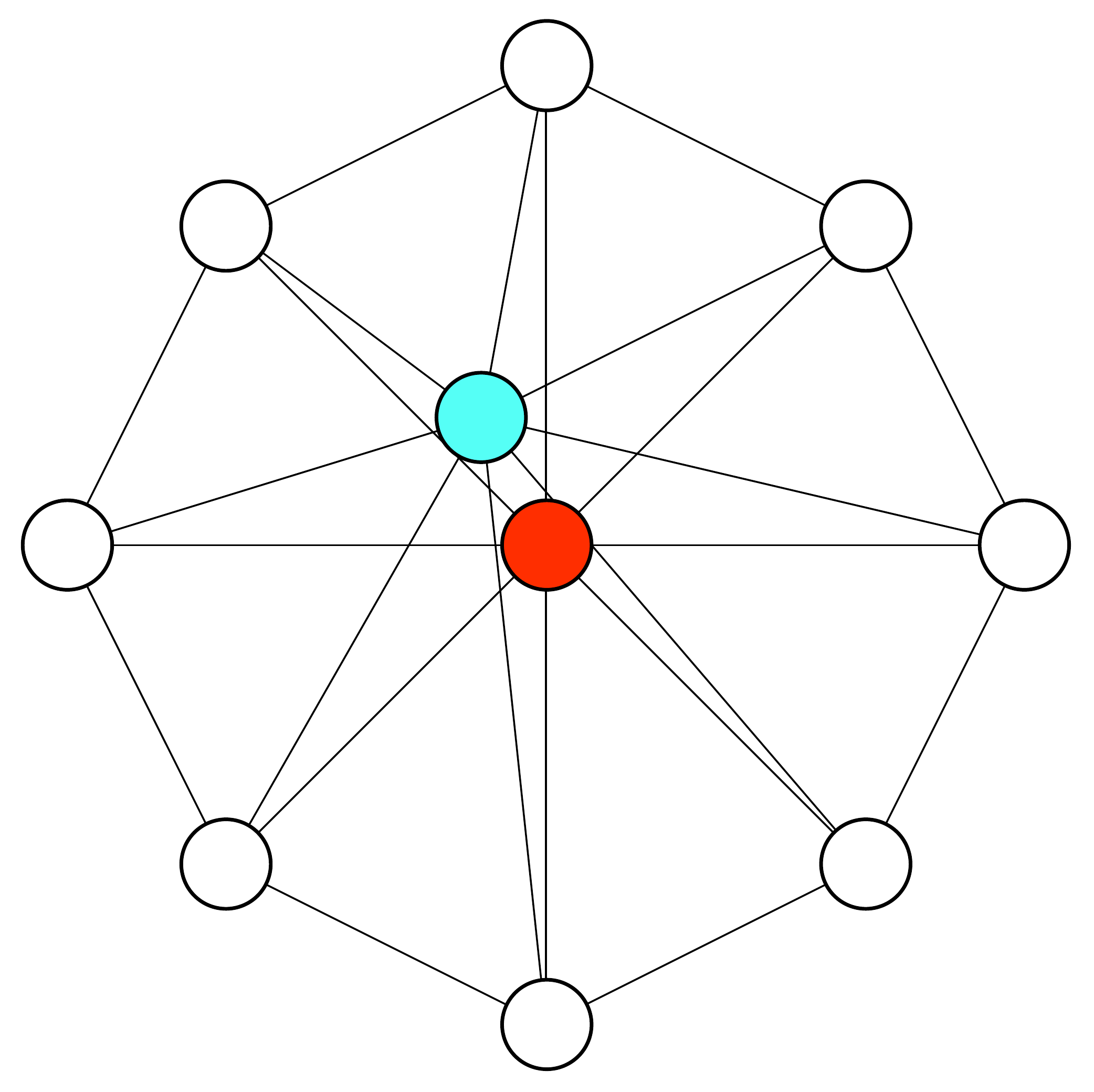}}\protect\caption{In these simple variations that retain the symmetry of the Lieb plaquette,
two localized states exist corresponding to two possible choices of
angular momentum. The size of the circles that represent lattice sites
indicates the relative value of the wavefunction amplitudes. There
is an analogy of these states with atomic orbitals. \label{fig:symmetry-2}}
\end{figure}

The previous argument for lattices for the Lieb's class can be generalized
to decorated lattices of the Mielke's and Tasaki's classes and other
decorated lattices using a symmetry argument. First, let us discuss
the case of the Lieb lattice (see Fig. \ref{fig:symmetry}). The tight-binding
Hamiltonian of one plaquette of the Lieb lattice has the symmetry
of a ring of 8 sites, that is, the plaquette Hamiltonian is invariant
in a $2\pi/8$ rotation of the set of site indices (or equivalently
in a circular permutation of the set of site indices). We emphasize
that this rotation should not be confused with a $2\pi/8$ rotation
in real space (the plaquette is not invariant in such a rotation).
However, the rotation of $2\pi/4$ in the set of sites indices can
be interpreted as a $2\pi/4$ rotation in real space. In the ring
of 8 sites, one has outer sites (that are connected to the rest of
the lattice) and inner sites (with connections only to sites of the
plaquette). In the case of Fig. \ref{fig:symmetry}, the inner sites
are indicated by the filled circles and the outer sites are given
by the empty circles. The generator of this rotation symmetry is the
equivalent of the angular momentum in the ring (note that in a 2D
lattice, the direction of the angular momentum is always perpendicular
to the lattice and therefore equal to $m\hbar$, where $m$ can be
interpreted as the $m$ in the ring momentum $k=m\cdot2\pi/N$) and
one can construct an eigenbasis of the Hamiltonian which is simultaneously
an eigenbasis of the angular momentum. The time reversal symmetry
of the Hamiltonian implies that each eigenstate of the plaquette tight-binding
Hamiltonian is degenerate with the respective state obtained in a
time reversal and these states have opposite angular momenta (this
is equivalent to stating that ring eigenstates with momenta $k$ and
$-k$ are degenerate). These two states can be added or subtracted,
generating the equivalent of the standing waves in the ring, that
is, states with zero probability density at certain sites of the cluster.
If the angular momentum is $\hbar N/4$, where $N$ is the number
of sites of the ring, one has zero probability density at the inner
sites or at the outer sites of the Lieb plaquette (nodes are separated
by $\lambda/2=2$). The latter will be a localized eigenstate not
only of the Lieb plaquette but also of the tight-binding Hamiltonian
of the full lattice. Note that this description is valid for any plaquettes
which have the same rotation symmetry as the ring. For example, one
could add additional sites at the center of the Lieb plaquette and
the rotation symmetry would remain, as shown in Fig. \ref{fig:symmetry-2}
(in all Figures, the relative size of the circles that represent lattice sites corresponds
to the relative value of the wavefunction amplitudes on the sites).

\begin{figure}
\subfloat[Rule I: symmetry.\label{fig:First-rule}]{\includegraphics[width=5cm]{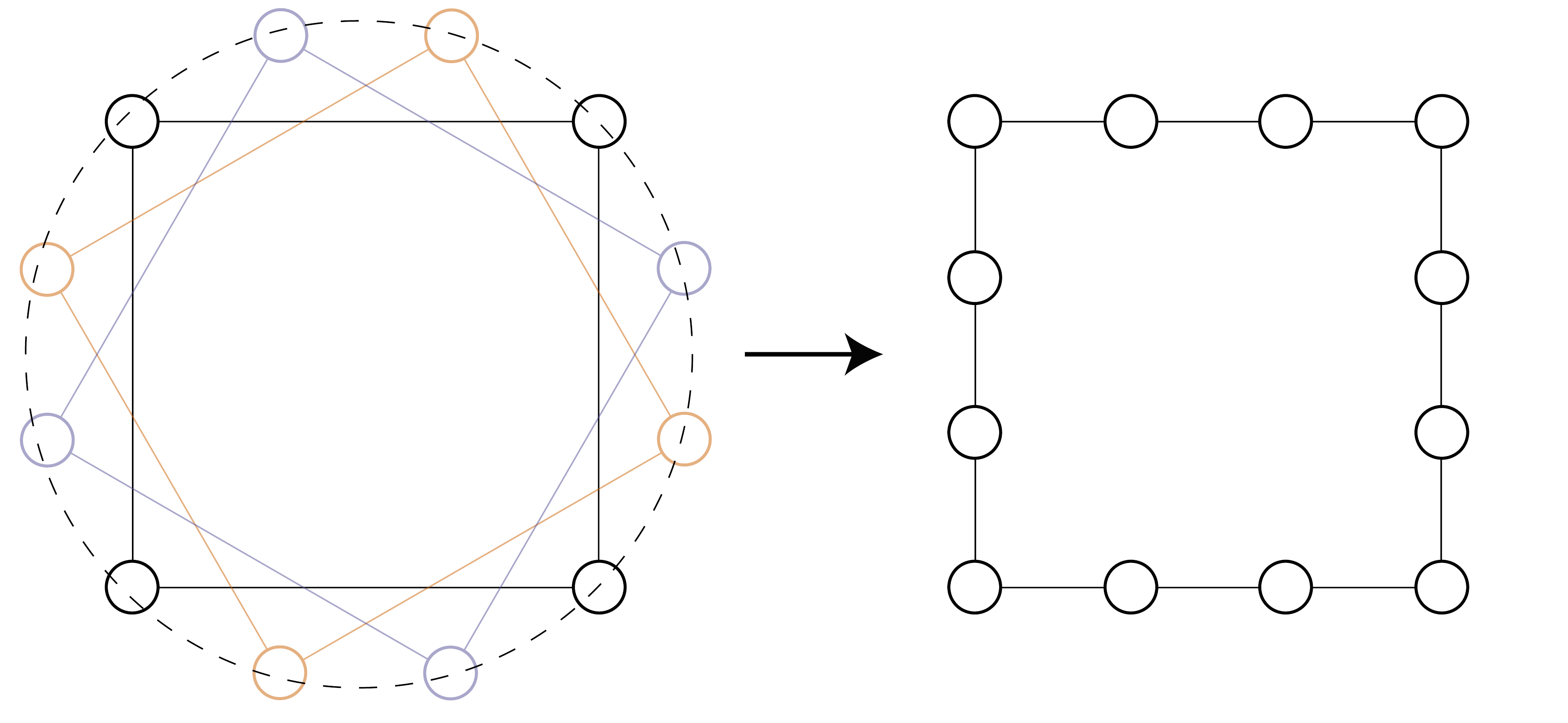}}

\subfloat[Rule II: add/remove nodes.\label{fig:second-rule}]{\includegraphics[width=6cm]{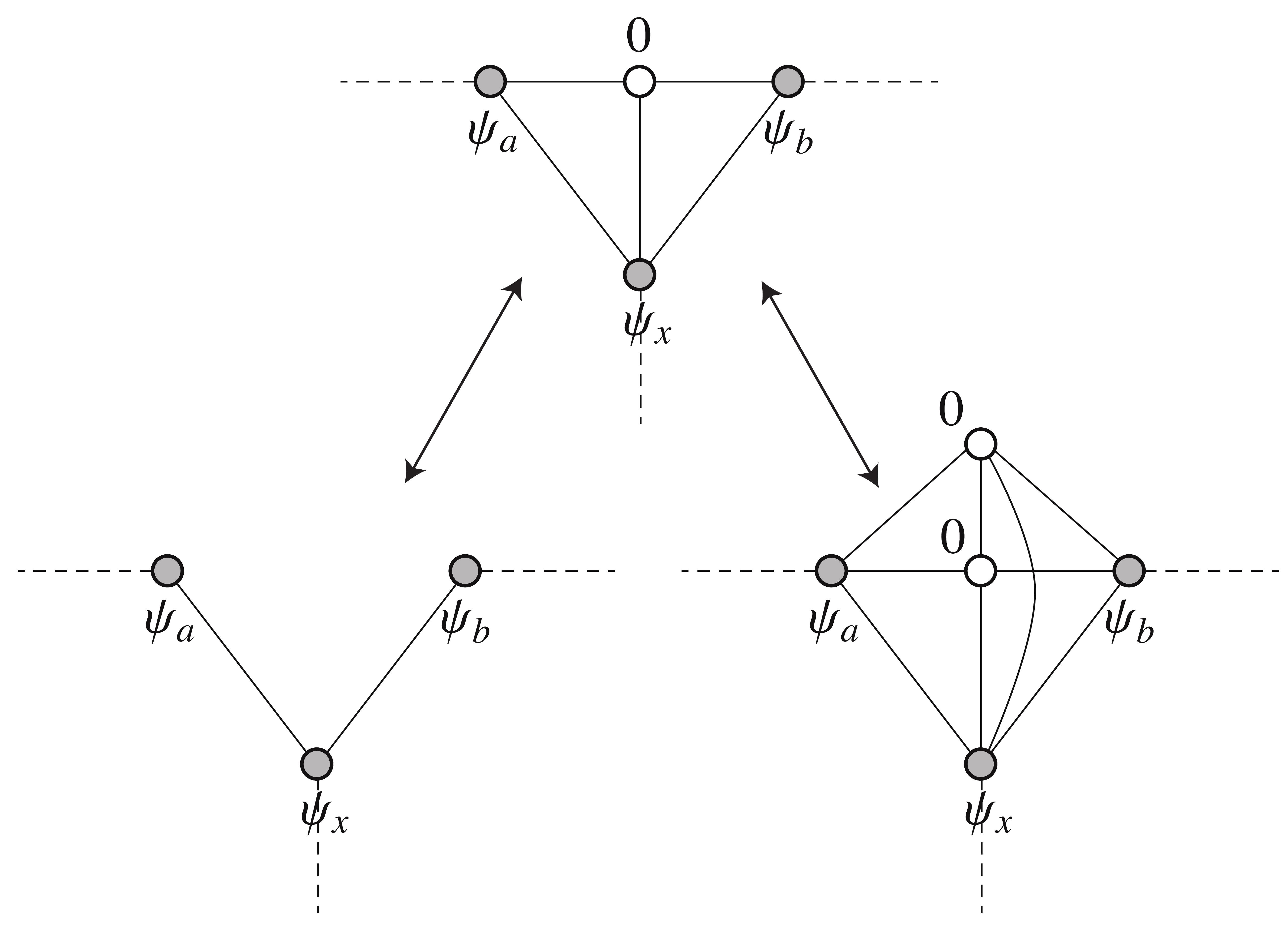}}

\subfloat[Rule III: folding ($\alpha=1/\sqrt{2}$).\label{fig:third-rule}]{\includegraphics[width=3.5cm]{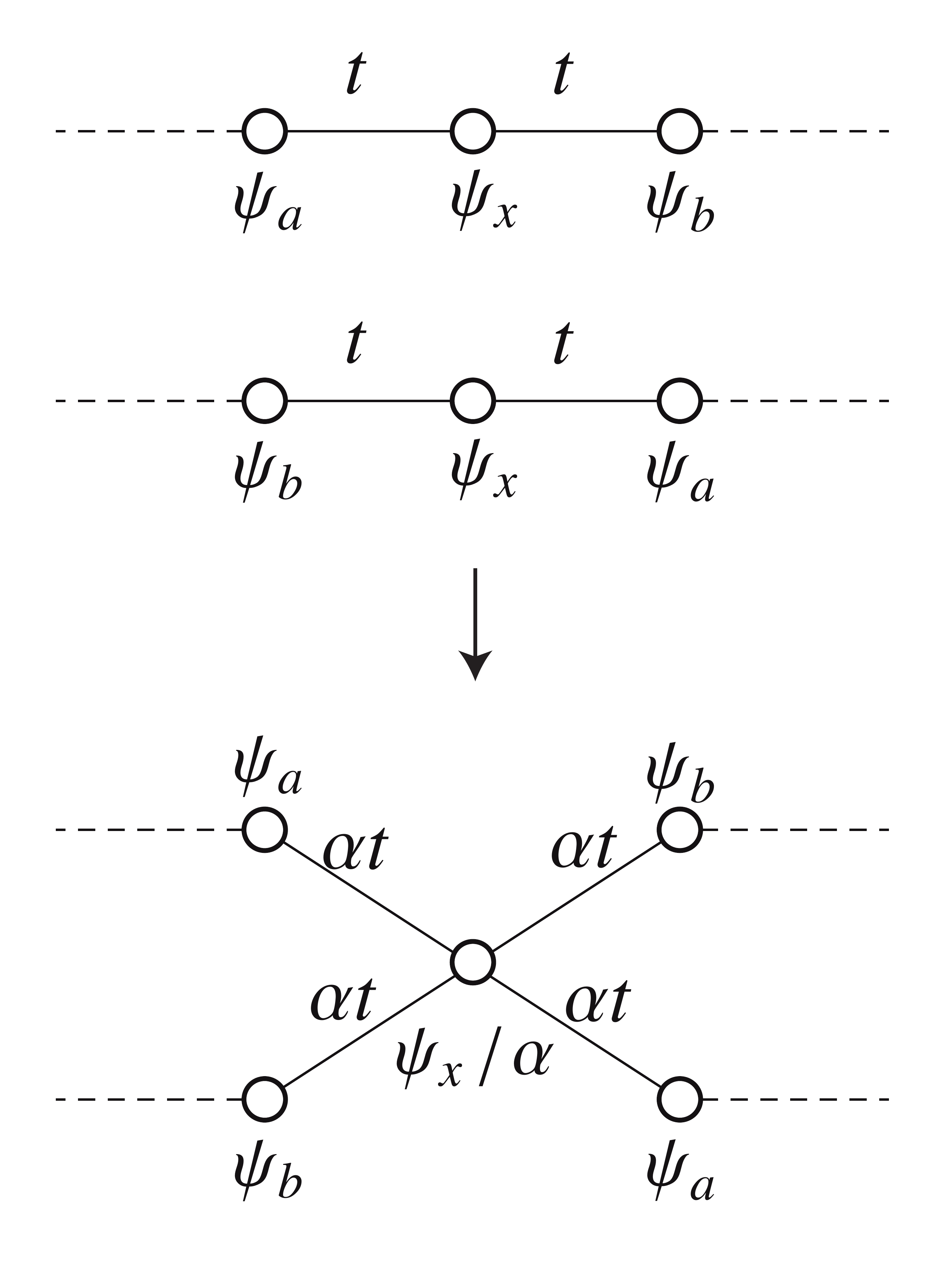}

}\subfloat[Rule IV: renormalization ($E=0$).\label{fig:fourth-rule}]{\includegraphics[width=4cm]{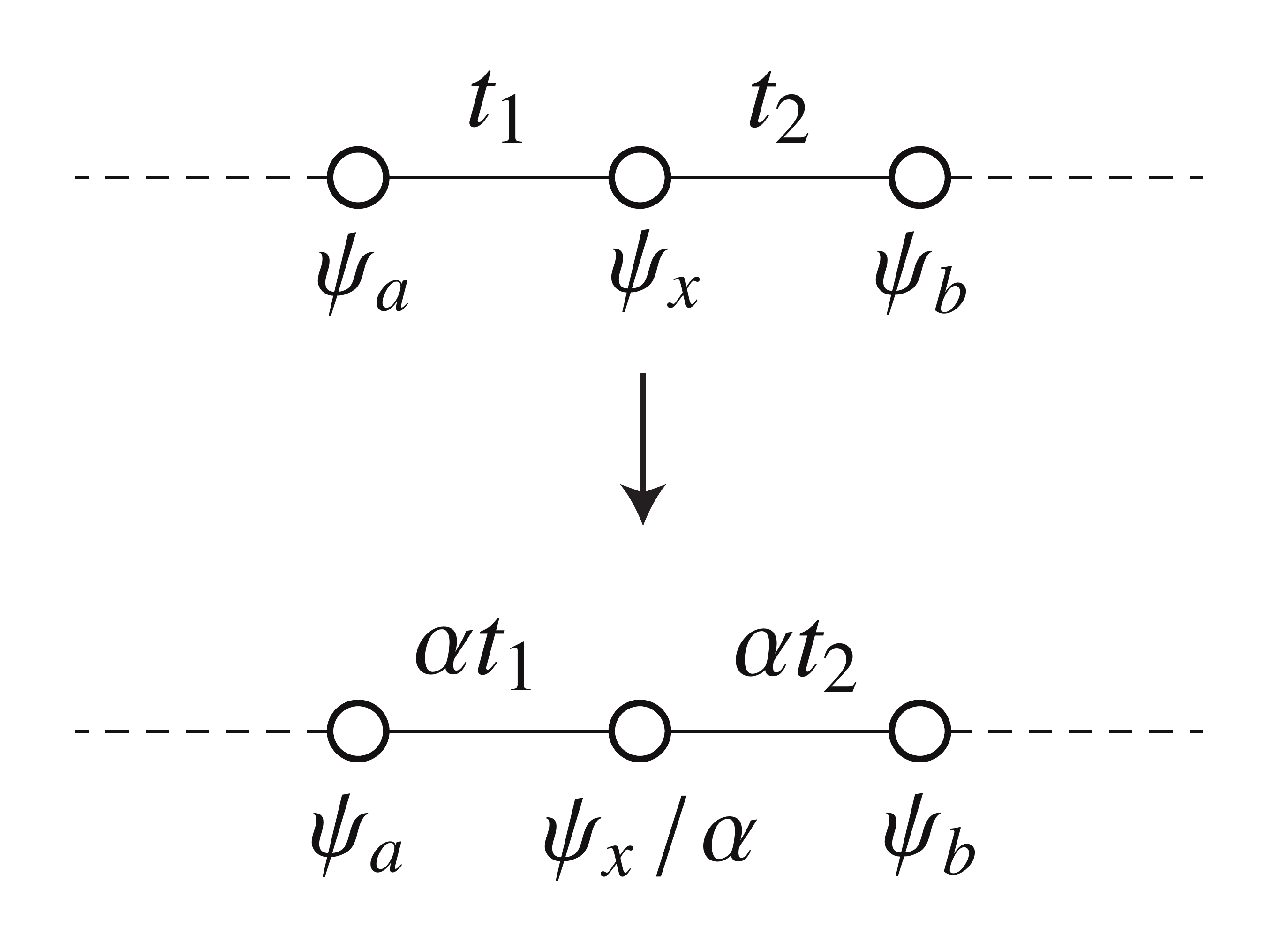}}

\subfloat[Rule V: unfolding ($\alpha=1/\sqrt{N}$).\label{fig:fifth-rule}]{\includegraphics[width=7cm]{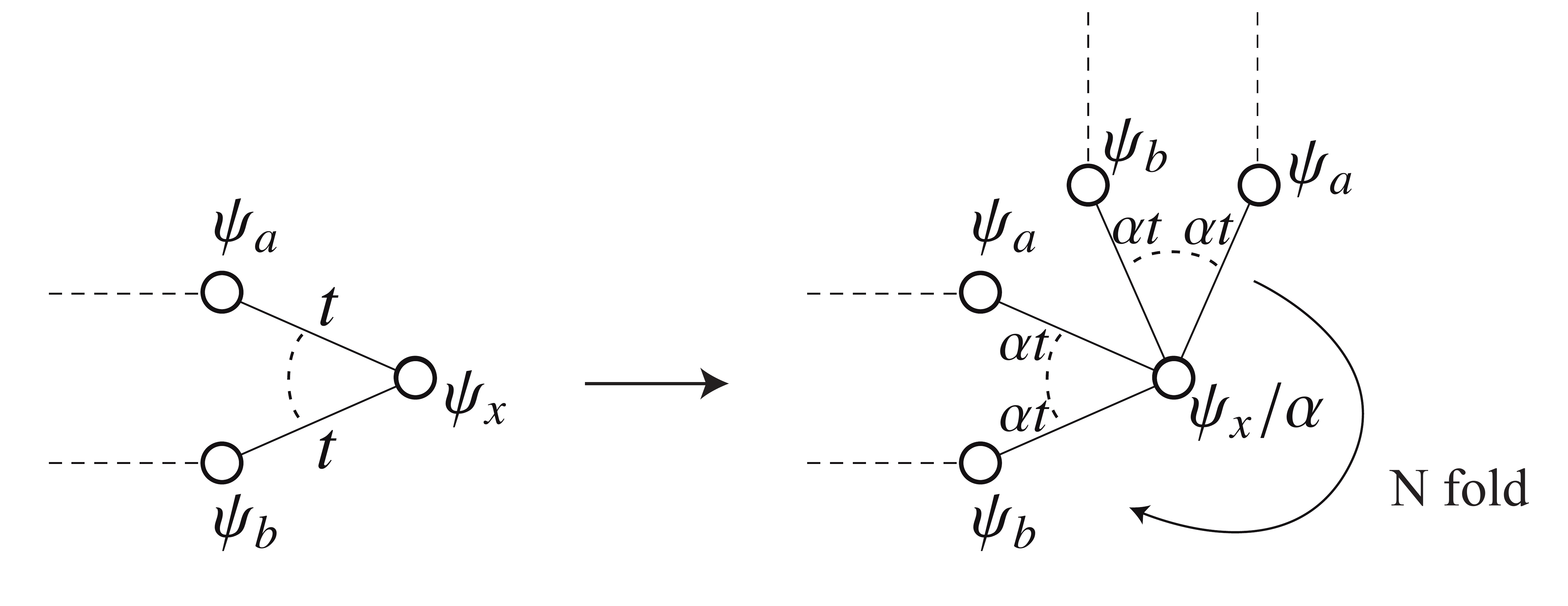}}\protect\caption{Set of origami rules for the construction of localized states in decorated
lattices. The values of $\alpha$ can be obtained from simple tight-binding
calculations, imposing the condition that the state is still an eigenstate
after applying the rule. \label{fig:rules}}
\end{figure}

Thus, our first rule is that localized states can be constructed if
a plaquette (or a set of adjacent plaquettes) has a larger symmetry
than the lattice, so that the Hamiltonian has two degenerate eigenstates
(which are simultaneously eigenstates of the generators of the symmetry
of the lattice) which have the same wavefunction values at the outer
sites of the plaquette (see Fig. \ref{fig:First-rule}). This rule
is enough to explain the existence of localized states in lattices
of the Lieb's class. More complex lattices with localized states can
be constructed by adding sites or hopping bonds that do not lower
the symmetry of the plaquette. These additional hoppings can be divided
into two sets: i) the set of hoppings from or to sites with probability
density nodes (these hoppings do not modify the energy of the
localized state); ii) the set of hoppings between
sites with finite density probability (these hoppings lower or raise
the energy of the localized state).

\begin{figure}
\subfloat[\label{fig:mielke}]{\includegraphics[width=4cm]{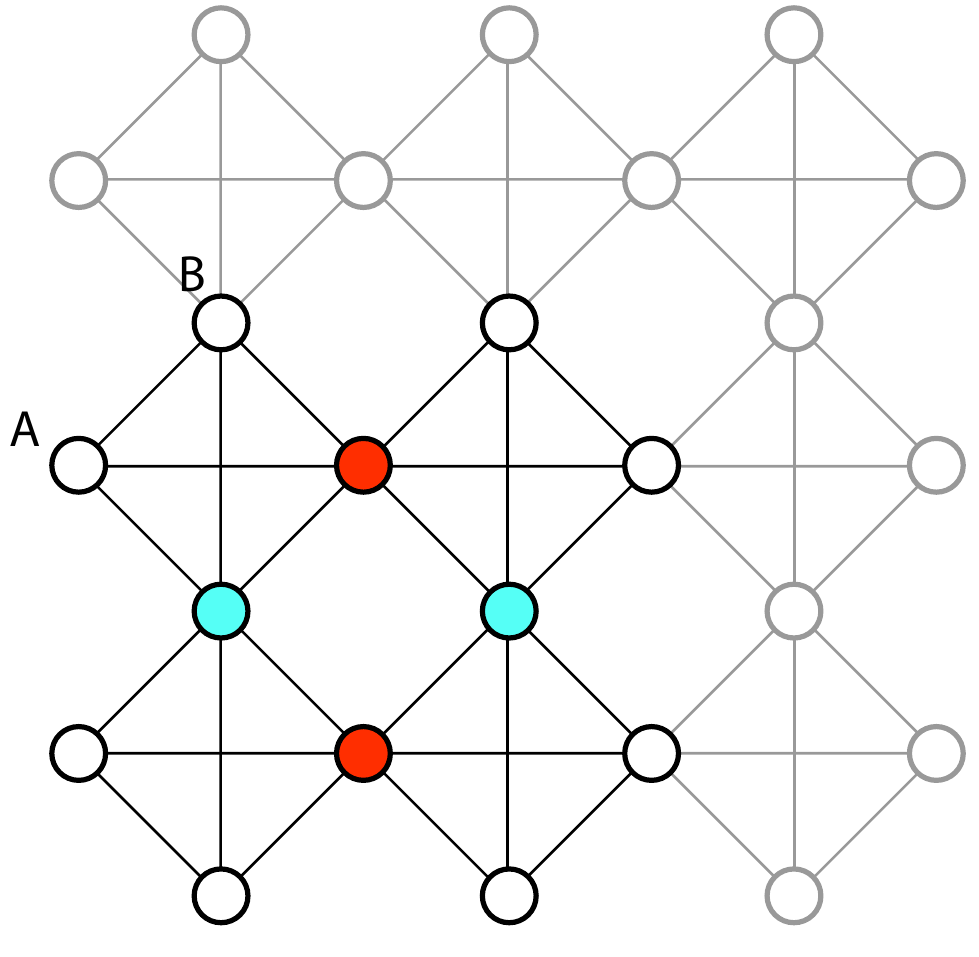}}\hfill{}\subfloat[\label{fig:tasaki}]{\includegraphics[width=4cm]{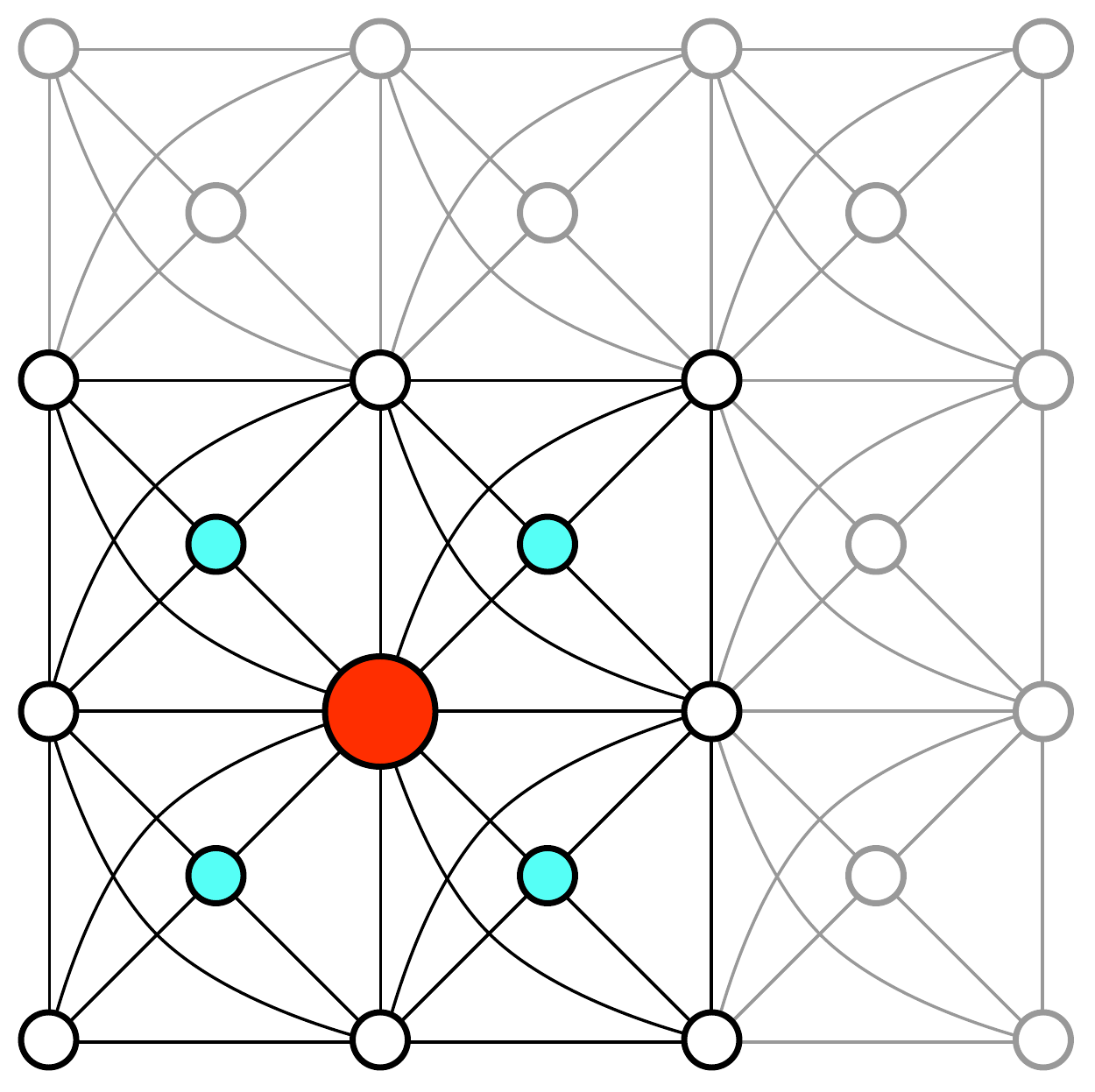}}\protect\caption{Localized states in the (a) Mielke lattice and (b) Tasaki lattice.
All hopping constants are the same ($t_{ij}=t$), except those associated
with curvy lines ($t_{ij}=t/2$). Note that the tight-binding Hamiltonian
for the Mielke plaquette is symmetric in the exchange of sites A and
B, but the full tight-binding Hamiltonian is not. This implies the
Hamiltonian eigenfunctions must have the same amplitude value (or
opposite values) at sites A and B. In the case of the localized states,
the value must be the same and the sites A and B are effectively one
site.\label{fig:symmetry-1}}
\end{figure}

A second rule for the construction of lattices with localized states
is the following. The existence of sites where a localized state has
probability density nodes does not affect the energy of the state
and these sites can be dropped, duplicated (as well as the respective
hopping bonds), or simply added (introducing appropriate hoppings
with neighboring sites) and the localized state remains an eigenstate
of the modified tight-binding model associated with the new plaquette
geometry (see Fig. \ref{fig:second-rule}). Furthermore, if one can
draw an axis through the plaquette that crosses only nodes, then dropping
these nodes one divides the localized state into two eigenstates of
the tight-binding Hamiltonians associated with the parts of the plaquette.
Bonds between nodes can also be dropped, added or duplicated. This
rule justifies the localized states in the lattice of Fig. \ref{fig:mielke}.
In fact, sites A and B in Fig. \ref{fig:mielke} can be seen as a
duplication of the equivalent site of the ring of Fig. \ref{fig:symmetry},
with the addition of a hopping bond between the duplicated nodes.

The third rule consists of the following: localized states can be
folded (adjusting the amplitude at the crossing and the respective
hoppings) along an axis that crosses the plaquette through sites that
have the same wavefunction values (see Fig. \ref{fig:third-rule});
if the folding is along an axis that crosses nodes, no adjustment
of hopping constants or wavefunctions amplitudes is needed.

The fourth rule is that the amplitude at a given site of a localized
state with zero energy can be renormalized without changing the energy
of the state, if the hopping constants to that site are renormalized
as well (see Fig. \ref{fig:fourth-rule}).

The fifth rule describes the unfolding of a plaquette around a given
site (see Fig. \ref{fig:fifth-rule}). Multiple rotated copies of
the original plaquette can be added around a site, provided that the
amplitude of the wavefunction on this site is adjusted, as well as
the hopping constants around this site.

\begin{figure*}
\includegraphics[width=0.9\textwidth]{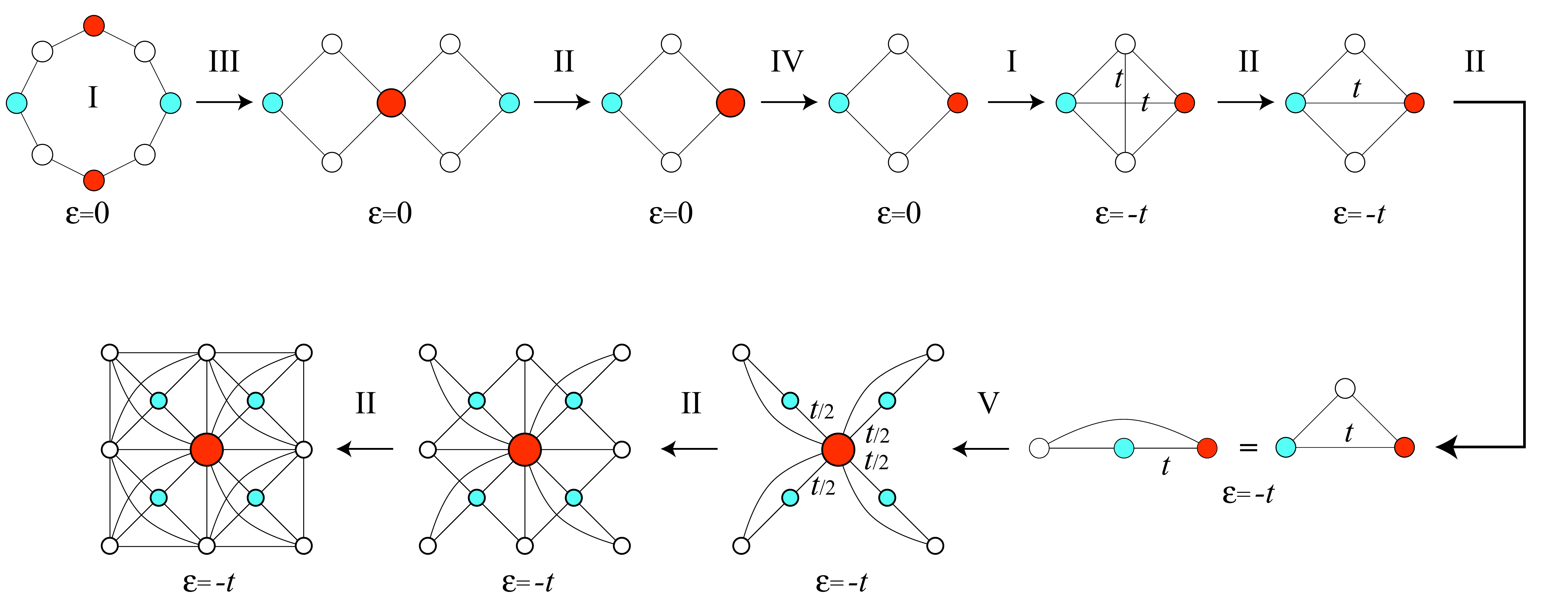}\protect\caption{Application of the rules presented in the text, starting from the
localized state in the Lieb plaquette and ending at the localized
state in the Tasaki lattice. Note that the energy of the localized
state in the Tasaki lattice is determined by the central hoppings
between sites with finite probability density.\label{fig:application}}
\end{figure*}

This set of rules justifies the existence of localized states in the
Mielke and Tasaki lattices of Fig. \ref{fig:symmetry-1}. In Fig.
\ref{fig:application}, we exemplify the application of this set of
rules starting from the localized state of the Lieb plaquette and
ending at the localized state of the Tasaki lattice.

We emphasize that these rules can be applied to construct localized states  in systems of arbitrary dimension, from 0D (a molecule) to 3D crystals, since the tight-binding bonds of Fig.~\ref{fig:rules} may not be coplanar and the unfolding axis can have an arbitrary direction.
Furthermore, the spin degree of freedom of a tight-binding  model may be interpreted as a  sublattice index, and spin flipping terms can be interpreted as hopping terms between such  sublattices. Localized states can also be created using the above origami rules involving the two (up and down spin) sublattices and examples of such localized states are the edge states in 1D models of topological insulators with spin flipping terms (see for example section 3.3. of Ref.~\onlinecite{Shen2013}).

Another context where this set of rules applies is that of one-magnon localized states in frustrated quantum Heisenberg antiferromagnets. In these systems, these states generate peculiar behavior such as magnetization plateaus around the saturation field \cite{Lacroix2011,Derzhko2015}.

Let us now discuss how much of this method can be applied in the $U\rightarrow\infty$
limit of the Hubbard model. We start by considering the Lieb lattice
and then we generalize our conclusions to more complex decorated lattices.
As discussed previously, the Lieb plaquette is a 8-site tight-binding
ring. The eigenfunctions of the Hubbard ring in the strong coupling
limit can be written as a tensorial product of the eigenfunctions
of a tight-binding model of independent spinless fermions (holes)
in the ring with $L$ sites and the eigenfunctions of an Heisenberg
model (with exchange constant $J=t^{2}/U$) in a reduced chain \citep{Ogata1990,Gebhard1997,Dias1992,Peres2000}.
The ring of spinless fermions is threaded by a fictitious magnetic
flux, $\phi=q_{s}$, generated by the spin configurations in the reduced
Heisenberg chain (where $q_{s}$ is the total spin momentum) and the
eigenvalues to order $t$ are given by
\begin{equation}
E(k_{1},\ldots,k_{N_{h}+N_{d}})=-2t\sum_{i=1}^{N_{h}}\cos\left(k_{i}-\frac{q_{s}}{L}\right),\label{eq:eigen-1}
\end{equation}
where $k_{i}=(2\pi/L)n_{i}$, $n_{i}=0,\ldots,L-1$ are the momenta
of the holes in the spinless ring. It is obvious that the rules for
the construction of tight-binding localized states discussed above
also apply to the case of one hole in a saturated ferromagnetic background.
What one also concludes from the solution of the Hubbard ring in the
strong coupling limit is that the same applies to the case of one
hole moving in a Lieb plaquette with arbitrary spin configuration
as long as the spin momentum is zero (since non-zero spin momentum
destroys the time reversal symmetry of the tight-binding model of
spinless fermions).

Does this apply to more complex plaquettes that share the rotation
symmetry of the Lieb plaquette? Taking the example shown in Fig. \ref{fig:localizedouteronly},
one sees that in the non-interacting case, a localized state is present
where the particle is confined to a 1D path. This leads one to suggest
that an equivalent localized state can be constructed for the hole
moving in the spin background, if we impose a $q_{s}=0$ spin momentum
for the spins configuration in the 1D path. However one should note
that in Fig. \ref{fig:localizedouteronly}, despite the electron probability
density being finite only in the outer ring, when the electron is
at the outer ring, it still hops to the center site, but summing over
all the hopping possibilities from the sites at the outer ring to
the center site, the result will be zero (destructive interference).
In the case of the hole moving in the spin background, the hops of
the hole from sites at the outer ring to the center site mix the spins
at the outer ring and at the center. In order for one to have destructive
interference at the center, the spin configuration must be a uniform
linear combination of all possible permutations of the set of spins
(given the number of up spins and down spins) \citep{Dias1992}. The
reason is the following: when an electron in the localized state of
Fig. \ref{fig:localizedouteronly} hops from a site of the outer ring
to the center site, it interferes destructively with the contributions
of hoppings from the other sites of the outer ring. In the case of
the hole, one has the different spin backgrounds and the hops of the
hole from the outer ring to the center should not apparently interfere
destructively. However, if one works with hole states such that, for
a given number of up and down spins in the plaquette, the spin configuration
is a uniform linear combination of all possible permutations of the
set of spins, then the spin configuration in the outer ring has $q_{s}=0$
spin momentum, that is, one can freely perform circular shifts of
the spin configuration in the outer ring. This implies that, assuming
a spin configuration which is a uniform linear combination of all
possible permutations of the set of spins, a hole jump from a site
A or B of the outer ring to the central site will generate the same
final state, independently of the initial site being A or B. Therefore,
we have the same localized state for one hole in the $U\rightarrow\infty$
limit of the Hubbard model as for one electron when $U=0$. This construction
of a localized state is valid for any plaquette corresponding to a
decorated lattice where a localized state of one tight-binding electron
exists.

Note that besides the localized state degeneracy associated with the
choice of the lattice plaquette   where the localized state sits,
there is a huge degeneracy associated with the possible choices of
number of up spins (or down spins) in the plaquette and in the rest
of the lattice. This degeneracy is lifted by the Heisenberg corrections
of order $t^{2}/U$, as in the Hubbard ring \citep{Ogata1990,Gebhard1997,Dias1992,Peres2000}.

Another important remark is that while the one-electron localized
states of the Mielke and Tasaki lattices are the ground states of
the respective tight-binding Hamiltonians, in the case of the $U\rightarrow\infty$
Hubbard model, the one-hole localized state is not the ground state,
for the choice of relative hopping constants of Fig. \ref{fig:symmetry-1}.
However, since the exact form of the hole probability density is known
as well as the hole wavefunction phase (as in the case of the one-electron
localized state), it is possible to tune the geometry, hopping constants
and interactions in order to lower the energy of the hole localized
state relatively to the other states, so that the energy of the localized
state approaches the energy of the ground state.

In conclusion, we have presented a simple set of rules for the construction
of localized states of the Hubbard model in nearly arbitrary decorated geometries,
in the tight-binding limit ($U=0$), and in the strong-coupling limit ($U\rightarrow\infty$). The first step in this method is the choice of a plaquette
or a set of plaquettes with a higher symmetry than that of the whole
lattice. In this plaquette, one has a localized state of the tight-binding
Hamiltonian of the full lattice (this state has probability density
nodes at the sites shared between the plaquette and the rest of the
lattice). Using a simple set of rules, the tight-binding localized
state in such plaquette can be divided, folded or unfolded to new
plaquette geometries. We have shown that this set of rules can also be applied in the $U\rightarrow\infty$
limit of the Hubbard model, for the construction of localized states
of one hole in the plaquette, assuming a spin configuration which
is a uniform linear combination of all possible permutations of the
set of spins in the plaquette.
Note that in every other plaquette, one may place a localized hole, so localized hole states exist for hole doping between zero and  a value of the order of $1/N_{\text{plaq}}$ (the value depends on the lattice geometry), where $N_{\text{plaq}}$ is the number of plaquette sites.

This paper presents a unifying picture of construction of localized states, in tight-binding systems of arbitrary dimension (from 0D to 3D), arbitrary geometry (including Mielke's and Tasaki's 2D geometries), without and with interactions ($U=0$ or $U=\infty$, extending in the latter case the filling intervals where localized states are known to occur). The existence of localized states due to spin flipping terms in tight-binding descriptions of topological models, or the existence of one-magnon localized states in frustrated Heisenberg antiferromagnets, are two other contexts included in this unifying picture.

\begin{acknowledgments}
\appendix
R. G. Dias acknowledges the financial support by FEDER funds through the COMPETE 2020 Programme and National Funds throught FCT - Portuguese Foundation for Science and Technology under the project UID/CTM/50025/2013.
J. D. Gouveia acknowledges the financial support from the Portuguese
Science and Technology Foundation (FCT) through the grant SFRH/BD/73057/2010.
\end{acknowledgments}

\bibliographystyle{unsrt}
\bibliography{prl}

\end{document}